\documentclass[conference]{IEEEtran}
\IEEEoverridecommandlockouts
\usepackage{xcolor}

\usepackage{adjustbox}
\usepackage{cite}
\usepackage{amsmath,amssymb,amsfonts}
\usepackage{amsthm}
\usepackage{graphicx}
\usepackage{textcomp}
\usepackage{xcolor}
\usepackage{algorithm}
\usepackage{array}
\usepackage{bm}
\usepackage{algorithm}
\usepackage{algpseudocode}
\usepackage{caption}
\usepackage{subcaption}
\def\BibTeX{{\rm B\kern-.05em{\sc i\kern-.025em b}\kern-.08em
    T\kern-.1667em\lower.7ex\hbox{E}\kern-.125emX}}
\usepackage{enumitem}
\usepackage[colorlinks]{hyperref}
\usepackage[dvipsnames]{xcolor}
\hypersetup{
    colorlinks = true,
    linkcolor  = blue,       
    citecolor  = blue,       
    urlcolor   = blue        
}
\usepackage{textcomp}
\usepackage{stfloats}
\usepackage{url}
\usepackage{verbatim}
\def\BibTeX{{\rm B\kern-.05em{\sc i\kern-.025em b}\kern-.08em
    T\kern-.1667em\lower.7ex\hbox{E}\kern-.125emX}}

    \usepackage{fancyhdr,lipsum}
 \fancypagestyle{mahmood}{%
    \fancyhf{} 
    
 }%
 \makeatletter
 \let\ps@IEEEtitlepagestyle\ps@mahmood
 \makeatother
\begin{document}


\title{
World Model for Energy Efficiency Maximization in Dynamic Metasurface Antennas 
}
\author{\IEEEauthorblockN{Abdolrasoul Sakhaei Gharagezlou, Mehdi Monemi, Matti Latva-aho, and Mehdi Rasti}
\IEEEauthorblockA{Centre for Wireless Communications (CWC),
University of Oulu, Oulu, Finland\\
Emails:  \{abdolrasoul.sakhaeigharagezlou; mehdi.monemi; matti.latva-aho; mehdi.rasti\}@oulu.fi}

}

\IEEEaftertitletext{\vspace{-2 em}}
\maketitle

\begin{abstract}
Dynamic metasurface antennas (DMAs) have emerged as a promising technology for realizing large-scale antenna arrays in next-generation wireless networks owing to their ability to enable dense and energy-efficient antenna architectures. However, jointly optimizing digital beamforming and DMA element weights under time-varying channel conditions remains challenging due to their coupled effects on the received signal and interference powers. We formulate this problem to maximize time-averaged energy efficiency (EE) under time-varying channel conditions, subject to transmit power constraints, Lorentzian phase limitations, and minimum rate requirements. To enable long-term planning and improve sample efficiency compared with model-free reinforcement learning, we develop a world model framework that learns latent system dynamics from interactions with the environment and jointly optimizes digital beamforming and DMA element weights through imagined trajectories. Furthermore, we incorporate feasibility-preserving action mappings and safety cost signals to ensure constraint satisfaction during policy learning. Simulation results demonstrate that the proposed framework improves the time-averaged EE by approximately 20\% compared with the deep Q-network baseline, while achieving higher sample efficiency and more reliable constraint satisfaction.
\end{abstract}


\section{Introduction}
Next-generation wireless networks are expected to rely on large antenna arrays at base stations (BSs) to achieve higher beamforming gains and improved connectivity. Dynamic metasurface antennas (DMAs) have emerged as a promising technology for implementing such arrays with reduced hardware cost and power consumption. In DMA architectures, multiple radiating metamaterial elements are arranged along microstrip waveguides, with each microstrip fed by a dedicated radio-frequency chain. However, jointly optimizing digital beamforming and DMA element weights remains challenging, particularly under time-varying channel conditions \cite{DMAZhang,frequnctflatDMA}.

Extensive research has focused on DMA-based antenna architectures to enhance wireless communication performance \cite{DMAZhang, frequnctflatDMA, Rasoulpower, RasoulRSMA}. For example, the authors in \cite{DMAZhang} investigated near-field beam focusing for sum-rate maximization in multi-user downlink systems considering three antenna architectures, namely fully digital, phase-shifter-based hybrid, and DMA. The authors in \cite{frequnctflatDMA} proposed a novel algorithm for maximizing the user's signal-to-noise ratio and derived a closed-form expression for the DMA weights. The authors in \cite{Rasoulpower} investigated spectral efficiency maximization in DMA-based systems under a realistic transmit-power constraint that accounts for the actual radiated power. The authors in \cite{RasoulRSMA} investigated rate-splitting multiple access for DMA-based wideband communications, considering the frequency-selective Lorentzian response of the metasurface elements to maximize the minimum sum rate across subcarriers. However, the effectiveness of the optimization-based methods employed in these studies largely depends on accurate mathematical models of the wireless network, which are often difficult to obtain in practice due to time-varying channel conditions and incomplete system information. Furthermore, these methods typically require iterative solutions to non-convex optimization problems at each time slot, imposing substantial computational complexity that limits their applicability to real-time operation \cite{optimizationLi}. Although heuristic algorithms offer alternative approaches to addressing non-convex optimization problems, they often exhibit limited scalability and robustness in large-scale and highly dynamic wireless networks \cite{huristic}.

To address these challenges, recent studies have explored reinforcement learning (RL) approaches, including model-based RL (MBRL) and model-free RL (MFRL), to optimize wireless network performance under time-varying channel conditions without requiring complete knowledge of the underlying system dynamics \cite{MBRL2,MFRL2}. For example, the authors in \cite{RLDMA} developed a deep RL framework to optimize frequency-selective DMA weights for maximizing the average beamforming gain across subcarriers in near-field uplink communications. However, existing RL-based approaches still face several challenges, including data inefficiency, which requires extensive environment interactions to explore high-dimensional action spaces, and limited long-term planning capabilities, which may hinder the joint optimization of digital beamforming and DMA element weights under time-varying channel conditions.

World-model-based RL (WMBRL) offers a promising approach to addressing these challenges by learning compact representations of system dynamics and optimizing policies through imagined trajectories. This enables sample-efficient policy optimization by leveraging predicted future states to jointly optimize digital beamforming and DMA element weights under time-varying channel conditions. {\it Such long-term planning is particularly important for maximizing time-averaged energy efficiency (EE) under time-varying channel conditions \cite{worldmodelGuangyuan}, as it enables the joint optimization of digital beamforming and DMA element weights across multiple time slots rather than independently optimizing them for each channel realization. Nevertheless, jointly optimizing digital beamforming and DMA element weights through long-term planning and safety-aware learning under time-varying channel conditions remains largely unexplored.}

\textit{Contributions:} Motivated by the above research gap, the primary contributions of this paper are outlined as follows:
\begin{itemize}
    \item We develop a unified optimization framework for jointly optimizing digital beamforming and DMA element weights to maximize time-averaged EE under time-varying channel conditions, subject to transmit power constraints, Lorentzian phase limitations, and minimum rate requirements.

    \item We propose a constraint-handling approach that enforces transmit power constraints and Lorentzian phase limitations through feasibility-preserving action mappings, while addressing minimum rate requirements through a safety-cost mechanism. This approach reduces the learning complexity while enabling explicit control over constraint satisfaction during policy optimization.

    \item We develop a WMBRL framework that learns system dynamics from collected experience and leverages imagined trajectories to jointly optimize digital beamforming and DMA element weights, enabling long-term planning while reducing the need for extensive interactions during policy learning. 
    
    \item We demonstrate through simulations that the proposed framework achieves an average improvement of approximately 20\% in time-averaged EE over the deep Q-network (DQN) baseline in the considered system, while exhibiting higher sample efficiency and more reliable constraint satisfaction.
\end{itemize}

\textit{Notation:} Throughout this paper, boldface lowercase and uppercase letters denote vectors and matrices, respectively. The superscript $(\cdot)^{H}$ represents the conjugate transpose, while $\|\cdot\|$ denotes the Euclidean norm. The operators $\mathbb{E}\{\cdot\}$ and  \(\mathbb{I}\{\cdot\}\) denote the statistical expectation and the indicator function, respectively. The operators $\Re\{\cdot\}$ and
$\Im\{\cdot\}$ denote the real and
imaginary parts, respectively. Furthermore, $\mathbb{C}^{M\times N}$ denotes the space of $M\times N$ complex-valued matrices, $\mathcal{CN}(\mu,\sigma^2)$ represents the circularly symmetric complex Gaussian distribution with mean $\mu$ and variance $\sigma^2$, and $\mathbf{I}_N$ denotes the $N\times N$ identity matrix.

\section{System Model and Problem Formulation}\label{Sec2}
We consider a DMA-based BS serving \(K\) single-antenna users. The BS is equipped with \(N=N^{\mathsf d}\times N^{\mathsf e}\) metamaterial elements arranged along \(N^{\mathsf d}\) microstrips, each comprising \(N^{\mathsf e}\) subwavelength radiating elements. The users are assumed to be randomly located around the BS, with horizontal distances within \([\psi^{\mathsf{min}},\psi^{\mathsf{max}}]\). The transmission period is divided into \(T\) time slots.

\subsection{Time-Varying Channel Model}
The channel vector between the BS and the \(k\)-th user at the \(t\)-th time slot is expressed as $\mathbf{g}_k[t] = \sqrt{\nu_k} \mathbf{\widetilde{g}}_k[t], \forall t, k$, where $\nu_k = \Big(\frac{c_0}{4\pi f^\mathsf c \psi_k}\Big)^2$ denotes the free-space channel power gain, $c_0$ is the speed of light, $f^\mathsf c$ indicates the carrier frequency, and $\psi_k \in [\psi^\mathsf{min}, \psi^\mathsf{max}]$ is the distance between the BS and $k$-th user. Furthermore, \(\widetilde{\mathbf g}_k[t]\) represents the small-scale fading vector. To capture the time-varying nature of the wireless channels, the small-scale fading components are modeled using a first-order complex Gauss–Markov process following the Jakes fading model, expressed as follows \cite{channel, worldmodelGuangyuan}:
\begin{equation}
\mathbf{\widetilde{g}}_k[t]=\rho \mathbf{\widetilde{g}}_k[t-1]+
\sqrt{1-\rho^2}\,\mathbf{e}_k[t], \quad \forall t,k,
\end{equation}
where \(\widetilde{\mathbf g}_k[0]\sim\mathcal{CN}(\mathbf 0,\mathbf I_N)\) denotes the initial small-scale fading vector, and \(\mathbf e_k[t]\sim\mathcal{CN}(\mathbf 0,\mathbf I_N)\) represents an independent innovation process. Furthermore, \(0\leq\rho<1\) denotes the temporal correlation coefficient between consecutive small-scale fading realizations.

\subsection{DMA Architecture and Signal Model}
 The frequency response of each element at the $t$-th time slot is formulated as follows \cite{frequnctflatDMA}:
\begin{equation}\label{frequencyflat}
    q_{i,\ell}[t]\in \mathcal{Q}_{i,\ell}[t]=\Big\lbrace \frac{j+e^{j\epsilon_{i,\ell}[t]}}{2} \Big\vert \epsilon_{i,\ell}[t]\in [0,2\pi] \Big \rbrace, \forall t,i,\ell,
\end{equation}
where $i\in\lbrace 1, \dots, N^\mathsf d\rbrace$ and $\ell \in\lbrace 1, \dots, N^\mathsf e\rbrace$. Furthermore, signal propagation along each microstrip is characterized by attenuation and phase shift, where attenuation depends on the physical properties and geometry of the microstrip, while the phase shift is determined by the propagation wavenumber and the position along the waveguide. Accordingly, the signal propagation along each microstrip is modeled as $m_{i,\ell} = e^{-\zeta_{i,\ell}(\alpha_i+j\beta_i)}, \forall i,\ell$, where \(\zeta_{i,\ell}\) represents the position of the \(\ell\)-th radiating element along the \(i\)-th microstrip, while \(\alpha_i\) and \(\beta_i\) denote the corresponding attenuation coefficient and propagation wavenumber, respectively \cite{DMAZhang}. We assume that all microstrips are of the same type; $\alpha_{i}= \alpha$ and $\beta_i=\beta$. The equivalent signal transmitted by DMA radiating elements at the \(t\)-th time slot is expressed as $\mathbf{x}[t] = \mathbf{M} \mathbf{Q}[t] \mathbf{v}[t], \forall t$, where \(\mathbf{M}\in\mathbb{C}^{N\times N}\) denotes a diagonal matrix with diagonal entries defined as \(\mathbf{M}_{(i-1)N^{\mathsf e}+\ell,(i-1)N^{\mathsf e}+\ell}=m_{i,\ell}\). Furthermore, \(\mathbf{Q}[t]\in\mathbb{C}^{N\times N^{\mathsf d}}\) represents the configurable DMA weight matrix at the \(t\)-th time slot, with its elements defined as follows \cite{DMAZhang}:
\begin{equation}\label{DMAweights}
\mathbf{Q}_{(i-1)N^{\mathsf e}+\ell,n}[t]=
\begin{cases}
q_{i,\ell}[t], & i=n,\\
0, & i\neq n.
\end{cases}
\end{equation}
Finally, \(\mathbf{v}[t]=\big[v_1[t],\ldots,v_{N^{\mathsf d}}[t]\big]^T\) represents the input signal vector to the DMA microstrips at the \(t\)-th time slot, expressed as $\mathbf{v}[t]=\sum_{k=1}^{K}\mathbf{w}_k[t]s_k[t], \forall t$, where $\mathbf{w}_k[t]\in \mathbb{C}^{N^{\mathsf d}\times 1}$ is the digital precoding vector for $s_k$ at time slot $t$. The transmitted symbols are assumed to be normalized and mutually uncorrelated, satisfying $\mathbb{E}\lbrace\left|s_k[t]\right|^2\rbrace=1$, $\forall t, k$, and $\mathbb{E}\lbrace s_k[t]s_j^*[t]\rbrace=0, k\neq j$. Therefore, the received signal at user $k$ during the $t$-th time slot is expressed as follows:
\begin{align}\label{recievedsignal}
    y_k[t]&=\mathbf{g}_k^H[t]\mathbf{M}\mathbf{Q}[t]\mathbf{w}_k[t]s_k[t]\nonumber \\ 
    &+\sum_{j=1,j\neq k}^{K} \mathbf{g}_k^H[t]\mathbf{M}\mathbf{Q}[t]\mathbf{w}_j[t]s_j[t]+ n_k[t], \, \forall t,k,
\end{align}
where $n_k[t] \sim \mathcal{CN}(0,\sigma_k^2)$ is the additive white Gaussian noise at user $k$ during the $t$-th time slot, with zero mean and variance $\sigma_k^2$. Based on \eqref{recievedsignal}, the achievable rate of user \(k\) at the \(t\)-th time slot is expressed as $\Gamma_k[t] = \log_2\left(1+ \Pi_k[t]\right)$, where $\Pi_k[t]=\frac{\vert\mathbf{g}_k^H[t]\mathbf{M}\mathbf{Q}[t]\mathbf{w}_k[t]\vert^2}{\sum_{j=1,j\neq k}^{K} \vert\mathbf{g}_k^H[t]\mathbf{M}\mathbf{Q}[t]\mathbf{w}_j[t]\vert^2+\sigma^2_k}$ denotes the signal-to-interference-plus-noise ratio of user $k$ at time slot $t$.
The EE of the system at the \(t\)-th time slot is expressed as follows:
\vspace{-5pt}
\begin{equation}
    \eta[t] = \frac{\sum_{k=1}^{K} \Gamma_k[t]}{P^{\mathsf{rad}}[t]+P^\mathsf c},    
\end{equation}
where $P^{\mathsf{rad}}[t] = \sum_{k=1}^{K}\left\|\mathbf{M}\mathbf{Q}[t]\mathbf{w}_k[t]\right\|^2$ represents the total radiated power of the DMA elements at time slot $t$, while $P^\mathsf c$ is the fixed circuit power \cite{Rasoulpower}.

\subsection{Problem Formulation}
The objective of this paper is to maximize the time-averaged EE over \(T\) time slots by jointly optimizing the digital beamforming vectors \(\{\mathbf{w}_k[t]\}\) and DMA weight matrix \(\mathbf{Q}[t]\) under time-varying channel conditions, subject to transmit power constraints, Lorentzian phase limitations, and minimum rate requirements for each user at each time slot. The resulting optimization problem is formulated as follows:
\vspace{-5pt}
\begin{subequations}\label{eq:problem_DMA}
\begin{align}
\mathcal{P}1:
\underset{\{\mathbf{w}_k[t],\mathbf{Q}[t]\}_{t=1}^{T}}{\operatorname{maximize}}\,\,\,
&\frac{1}{T}
\sum_{t=1}^{T}\eta[t]
\label{eq:objective_DMA}\\
\mathrm{s.t.}
&\sum_{k=1}^{K}
\left\|\mathbf{M}\mathbf{Q}[t]\mathbf{w}_k[t]\right\|^2\!+\!P^\mathsf c\! 
\leq \!P^{\mathsf{max}},
\label{eq:power_DMA}\\
&q_{i,\ell}[t]\in \mathcal{Q}_{i,\ell}[t], \quad \forall t,i,\ell, \label{eq:frequncyresponse}\\
& \Gamma_k[t]\ge \Gamma^{\mathsf{min}}, \quad \forall t,k.
\label{eq:SINR_DMA}
\end{align}
\end{subequations}
In problem $\mathcal{P}1$, constraint \eqref{eq:power_DMA} limits the total power consumption, including the radiated and circuit powers, to the maximum power budget $P^{\mathsf{max}}$. Constraint \eqref{eq:frequncyresponse} ensures that the DMA weights satisfy the Lorentzian phase response of the metamaterial elements, while constraint \eqref{eq:SINR_DMA} guarantees that the achievable rate of each user meets its minimum requirement $\Gamma^{\mathsf{min}}$ at each time slot.

\section{Proposed Solution}\label{secIII}
Problem \(\mathcal{P}1\) is a non-convex stochastic optimization problem due to time-varying channel conditions and the coupling between digital beamforming and DMA element weight variables. To address these challenges, we develop a WMBRL framework that learns latent system dynamics from collected experience and jointly optimizes digital beamforming and DMA element weights through imagined trajectories, enabling long-term planning while reducing the need for extensive environment interactions. Furthermore, the proposed framework handles operational constraints through feasibility-preserving action mappings and safety cost signals during policy learning.

\subsection{Observation, Action, and Reward Formulation}
To apply the WMBRL framework to problem $\mathcal{P}1$, we characterize the interactions between the agent and the wireless environment using observation, action, and reward representations, which are defined as follows:

\subsubsection{Observation Space}
At the beginning of each time slot \(t\), the agent observes the wireless environment through the following observation vector:
\vspace{-5pt}
\begin{align}\label{statespace1}
    \mathbf{o}[t] = \Big[&\Im\{\mathbf{g}_k[t]\}, \Re\{\mathbf{g}_k[t]\}, \Vert\mathbf{g}_k[t]\Vert^2,\nonumber
    \\ &\Pi_k[t-1], \Gamma_k[t-1], P^{\mathsf{rad}}[t-1]\Big],
\end{align}
where \(\Vert\mathbf{g}_k[t]\Vert^2\) represents the channel power of the \(k\)-th user. The observation-space dimension is \(d^{\mathsf o}=2NK+3K+1\).

\subsubsection{Action Space}
At each time slot \(t\), the action vector is defined as follows:
\vspace{-5pt}
\begin{equation}
\mathbf{a}[t]\!=\!\big[\bar{\boldsymbol{\epsilon}}[t],\{\bar{P}_k[t]\}_{k=1}^{K},\bar{P}_0[t]\big]\in[-1,1]^{d^{\mathsf a}},    
\end{equation}
where $\bar{\boldsymbol{\epsilon}}[t]=\big[\bar{\epsilon}_{i,\ell}[t]\big]_{i=1,\ell=1}^{N^\mathsf d,N^\mathsf e}$ controls the DMA element weights, \(\bar{P}_k[t]\) controls the transmit power allocated to user \(k\), $\bar{P}_0[t]$ controls the total radiated power, and \(d^{\mathsf a}=N+K+1\) is the action-space dimension. The corresponding digital beamforming vector is constructed as \(\mathbf w_k[t]=\sqrt{P_k[t]}\mathbf l_k[t]\), where $P_k[t]$ denotes transmit power allocated to user $k$ and \(\mathbf l_k[t]\) denotes  the unit-norm regularized zero-forcing beamforming direction of user \(k\), given by $\mathbf l_k[t] = \frac{\bar{\mathbf l}_k[t]}{\Vert \mathbf{M}\mathbf{Q}[t]\bar{\mathbf l}_k[t]\Vert}$, in which \(\bar{\mathbf l}_k[t]\) is the \(k\)-th column of \(\boldsymbol{\Omega}^H[t](\boldsymbol{\Omega}[t]\boldsymbol{\Omega}^H[t]+\delta\mathbf I_K)^{-1}\), with \(\boldsymbol{\Omega}[t]=\mathbf G^H[t]\mathbf M\mathbf Q[t]\), \(\mathbf G[t]=[\mathbf g_1[t],\ldots,\mathbf g_K[t]]\), and \(\delta>0\) denoting a small regularization parameter. This normalization ensures \(\|\mathbf M\mathbf Q[t]\mathbf l_k[t]\|=1\), yielding \(P^{\mathsf{rad}}[t]=\sum_{k=1}^{K}P_k[t]\). 

\subsubsection{Reward Function}
The reward function aims to maximize the time-averaged EE, with the immediate reward at time slot \(t\) defined as \(r[t]=\frac{\sum_{k=1}^{K}\Gamma_k[t]}{P^{\mathsf{rad}}[t]+P^{\mathsf c}}\). Accordingly, the expected discounted cumulative reward over \(T\) time slots is expressed as follows:
\vspace{-5pt}
\begin{equation}
    J^{\mathsf r}(\pi)=\mathbb{E}_{\pi}\left[\sum_{t=1}^{T}\gamma^{t-1}r[t]\right],
\end{equation}
where \(\gamma\) denotes the discount factor and \(\pi\) represents the policy. Maximizing this objective encourages long-term EE improvement through the joint optimization of digital beamforming and DMA element weights over the transmission horizon.

\subsection{Constraint Handling}
The constraints in problem \(\mathcal{P}1\) are handled using different approaches based on their characteristics. Specifically, the transmit power and Lorentzian phase constraints are enforced through feasibility-preserving action mappings, while the minimum rate requirements are addressed using a safety-cost function during policy learning. These mechanisms are detailed as follows.

\subsubsection{Transmit Power Constraint}
To satisfy the transmit-power constraint, the actor determines the total radiated power and its allocation among users. Specifically, the power allocated to each user is obtained through the softmax function as $P_k[t]=P^{\mathsf{a}}[t]
\frac{\exp(\bar P_k[t])}
{\sum_{j=1}^{K}\exp(\bar P_j[t])}$, where \(P^{\mathsf{a}}[t] = P^\mathsf m (\frac{P^{\mathsf{max}}-P^\mathsf c}{P^\mathsf m})^{\frac{\bar{P}_0[t]+1}{2}}\) denotes the total allocated radiated power, with $P^\mathsf m$ is the minimum radiated power. This transformation ensures that \(P_k[t]>0\) and \(\sum_{k=1}^{K}P_k[t]=P^{\mathsf{a}}[t]\le P^{\max}-P^{\mathsf c}\). Consequently, the transmit-power constraint in \eqref{eq:power_DMA} is satisfied at each time slot.

\subsubsection{Lorentzian Phase Constraint}
For each DMA element, the actor generates a normalized phase-control action $\bar{\epsilon}_{i,\ell}[t]\in[-1,1]$, which is mapped to the feasible phase interval as $\epsilon_{i,\ell}[t] = \pi(\bar{\epsilon}_{i,\ell}[t]+1)$. This mapping guarantees that $0\le \epsilon_{i,\ell}[t]\le 2\pi$. The corresponding DMA element weights are then obtained from the Lorentzian response in \eqref{frequencyflat} and arranged into the DMA weight matrix $\mathbf{Q}[t]$ according to \eqref{DMAweights}.

\subsubsection{Minimum Rate Requirements Constraint}
To incorporate the minimum-rate requirements into policy learning, we define a safety-cost function that quantifies rate constraint violations at each time slot as $c[t]=\frac{1}{K}\sum_{k=1}^{K}\mathbb{I}\{\Gamma_k[t]<\Gamma^{\mathsf{min}}\}$. The expected discounted cumulative safety cost is defined as $J^{\mathsf c}(\pi)=\mathbb E_{\pi}\left[\sum_{t=1}^{T}\gamma^{t-1}c[t]\right]$. Accordingly, the constrained policy optimization problem is formulated as follows:
\begin{equation}\label{LearningProb}
    \begin{aligned}
        \mathcal{P}2: \underset{\pi}{\operatorname{maximize}} \quad  J^\mathsf r(\pi) \,\,\,\,\operatorname{s.t.}\,\, J^\mathsf c(\pi) \leq b,
    \end{aligned}
\end{equation}
where \(b\) denotes the predefined safety-cost budget. This formulation encourages the agent to maximize time-averaged EE while reducing violations of the minimum-rate constraints. To solve the constrained policy optimization problem \(\mathcal{P}2\), we develop a WMBRL framework, as detailed in the following subsections.

\subsection{Learning Method}
The proposed WMBRL framework comprises three key components: (i) a \textit{\textbf{world model}} that learns latent representations of the system dynamics from observations and predicts future states, rewards, and safety costs; (ii) \textit{\textbf{reward and cost critics}} that evaluate the expected long-term rewards and safety costs associated with the learned policy; and (iii) a \textit{\textbf{safe actor}} that optimizes the policy to maximize the expected cumulative reward while satisfying safety constraints. The details of each component are presented in the following subsections.

\subsubsection{World Model Learning}
The world model incorporates an observation encoder, a recurrent state-space model (RSSM) to learn latent system dynamics, and prediction heads for reconstructing observations and predicting rewards and safety-cost signals. The observation encoder maps the observation \(\mathbf{o}[t]\) into a compact latent representation, while the RSSM captures the temporal evolution of the system through deterministic and stochastic latent states. Specifically, the deterministic state $\mathbf h[t]$ is updated using the previous deterministic state $\mathbf h[t-1]$, stochastic state $\mathbf z[t-1]$, and action $\mathbf a[t-1]$. The stochastic state $\mathbf z[t]$ is then inferred from the current observation features and deterministic state. The RSSM employs a transition prior to predict future latent states, enabling imagined trajectory generation without further environment interactions. The learned latent states enable observation reconstruction and the prediction of rewards and safety-cost signals. The corresponding world-model components are expressed as follows:
\begin{equation*}
\begin{aligned}
\mathrm{RSSM}
&\left\{
\begin{aligned}
&\text{Sequence model:}\,
\mathbf{h}[t]
\!=\!\!f_{\phi}\!\left(
\mathbf{h}[t\!-\!1],\mathbf{z}[t\!-\!1],\mathbf{a}[t\!-\!1]
\right),\\
&\text{Encoder:}\,
\mathbf{z}[t]
\sim q_{\phi}\!\left(
\mathbf{z}[t]\mid \mathbf{h}[t],\mathbf{o}[t]
\right),\\
&\text{Transition prior:}\quad
\hat{\mathbf{z}}[t]
\sim p_{\phi}\!\left(
\hat{\mathbf{z}}[t]\mid \mathbf{h}[t]
\right),
\end{aligned}
\right.
\\[1mm]
&\text{Reward predictor:}\,
\hat{r}[t]\sim p_{\phi}\!\left(
\hat{r}[t]\mid \mathbf{h}[t],\mathbf{z}[t]
\right),\\
&\text{Cost predictor:}\,
\hat{c}[t]\sim p_{\phi}\!\left(
\hat{c}[t]\mid \mathbf{h}[t],\mathbf{z}[t]
\right),\\
&\text{Decoder:}\quad
\hat{\mathbf{o}}[t]\sim p_{\phi}\!\left(
\hat{\mathbf{o}}[t]\mid \mathbf{h}[t],\mathbf{z}[t]
\right),
\end{aligned}
\end{equation*}
where \(\phi\) represents the trainable parameters of the world model, \(\hat{\mathbf z}[t]\) denotes the predicted latent state, and \(\hat r[t]\), \(\hat c[t]\), and \(\hat{\mathbf o}[t]\) represent the predicted reward, safety cost signal, and reconstructed observation, respectively. Furthermore, the functions \(q_\phi(\cdot)\) and \(p_\phi(\cdot)\) denote the posterior and predictive distributions, while \(f_\phi(\cdot)\) represents the deterministic state-transition function. The overall loss function of the world model is formulated as follows \cite{safedreamer}:
\vspace{-5pt}
\begin{equation*}\label{total-loss}
\mathcal{L}(\!\phi\!)\!\!=\!\mathbb{E}_{q_{\phi}}\!\!\left[\!\sum_{t=1}^{T}\!\!\left(\beta^{\mathsf{pred}} \mathcal{L}^{\mathsf{pred}}[t]\!+\!\beta^{\mathsf{tran}} \mathcal{L}^{\mathsf{tran}}[t]\!+\!\beta^{\mathsf{rep}}\mathcal{L}^{\mathsf{rep}}[t]\right)\!\!\right]\!,
\end{equation*}
where
\begin{equation*}
\begin{aligned}
\mathcal{L}^{\mathsf{pred}}[t]
=&-\log p_{\phi}\!\left(
\hat{\mathbf{o}}[t]\mid\mathbf{h}[t],\mathbf{z}[t]
\right) -\log p_{\phi}\!\left(
\hat{r}[t]\mid\mathbf{h}[t],\mathbf{z}[t]
\right) \\
&-\log p_{\phi}\!\left(
\hat{c}[t]\mid\mathbf{h}[t],\mathbf{z}[t]
\right),
\end{aligned}
\label{eq:prediction_loss}
\end{equation*}
\vspace{-10pt}
\begin{equation*}
\mathcal{L}^{\mathsf{tran}}[t]\!=\!\operatorname{max}\!\left\{\!
\!1,\operatorname{KL} \!\left[\operatorname{sg}\!\left(q_{\phi}\!\left(\mathbf{z}[t]\!\mid\!\mathbf{h}[t],\mathbf{o}[t]\right)\!\right)\!\Big\|p_{\phi}\!\left(\mathbf{z}[t]\!\mid\!\mathbf{h}[t]\right)\right]\!\right\},
\label{eq:dynamics_loss}
\end{equation*}
\vspace{-10pt}
\begin{equation*}
\mathcal{L}^{\mathsf{rep}}[t]\!\!=\!\max\!\left\{\!1,{\operatorname{KL}}\!\left[q_{\phi}\!\left(\mathbf{z}[t]\!\!\mid\!\!\mathbf{h}[t],\mathbf{o}[t]\right)\!\Big\|\!\operatorname{sg}\!\left(p_{\phi}\!\left(\mathbf{z}[t]\!\mid\!\mathbf{h}[t]\right)\right)\right]\!\right\}.
\label{eq:representation_loss}
\end{equation*}
with corresponding weighting coefficients \(\beta^{\mathsf{pred}}=1\), \(\beta^{\mathsf{tran}}=0.5\), and \(\beta^{\mathsf{rep}}=0.1\), while \(\operatorname{KL}(\cdot\|\cdot)\) denotes the Kullback–Leibler (KL) divergence between two probability distributions and $\operatorname{sg}(\cdot)$ is the stop-gradient operator. The prediction loss $\mathcal{L}^{\mathsf{pred}}$ jointly optimizes the observation decoder and the reward and safety-cost predictors. The transition loss $\mathcal{L}^{\mathsf{tran}}$ minimizes the  KL divergence between the predicted prior and posterior latent distributions, while the representation loss $\mathcal{L}^{\mathsf{rep}}$ minimizes the KL divergence between the posterior and predicted prior distributions to train the posterior encoder. These loss functions enable the world model to learn compact latent representations of system dynamics and predict future states, rewards, and safety costs to support policy optimization through imagined trajectories.

\subsubsection{Critic Learning}
The proposed framework incorporates two critic networks to estimate the expected long-term EE and safety costs of the learned policy using imagined trajectories predicted by the world model. Specifically, the reward critic estimates the expected cumulative reward, whereas the cost critic predicts the expected cumulative safety cost arising from constraint violations. Both critics are trained using temporal-difference \(\mathrm{TD}(\lambda)\) returns, which combine predicted immediate rewards or costs with estimates of future returns. Accordingly, the reward \(\mathrm{TD}(\lambda)\) return is recursively defined as follows \cite{safedreamerV3}:
\vspace{-5pt}
\begin{equation*}\label{eq:reward_td}
V_{\lambda}^{\mathsf r}[t]\!\!=\!\!
\begin{cases}
\!\hat r[t]\!\!+\!\gamma\left[\!
(1\!\!-\!\!\lambda)v_{\psi^{\mathsf r}}\!
\left(\mathbf h[t\!\!+\!1],\mathbf z[t\!\!+\!1]\!\right)\!
+\!\!\lambda V_{\lambda}^{\mathsf r}[t\!+\!1]
\right]\!, &\!\!\!\!\!\! t\!<\!\widetilde{H},\\[2mm]
v_{\psi^{\mathsf r}}
\left(\mathbf h[\widetilde{H}],\mathbf z[\widetilde{H}]\right), &\!\!\!\!\!\!\! t\!=\!\widetilde{H},
\end{cases}
\end{equation*}
where $v_{\psi^{\mathsf r}}(\cdot)$ denotes the reward critic with trainable parameters $\psi^{\mathsf r}$, $\widetilde{H}$ represents the imagination horizon, and $\lambda$ is the \(\mathrm{TD}(\lambda)\) parameter. Similarly, the cost \(\mathrm{TD}(\lambda)\) return is formulated as follows:
\vspace{-5pt}
\begin{equation*}\label{eq:cost_td}
V_{\lambda}^{\mathsf c}[t]\!=\!\!
\begin{cases}
\!\!\hat c[t]\!+\!\gamma\left[\!
(1\!\!-\!\!\lambda)v_{\psi^{\mathsf c}}
\!\left(\mathbf h[t\!\!+\!\!1],\mathbf z[t\!\!+\!\!1]\right)\!\!
+\!\!\lambda V_{\lambda}^{\mathsf c}[t\!+\!1]
\right]\!, & \!\!\!\!\!\!t\!<\!\!\widetilde{H}\!,\\[2mm]
v_{\psi^{\mathsf c}}
\left(\mathbf h[\widetilde{H}],\mathbf z[\widetilde{H}]\right), & \!\!\!\!\!\!\!t\!=\!\!\widetilde{H}\!,
\end{cases}
\end{equation*}
where \(v_{\psi^{\mathsf c}}(\cdot)\) denotes the cost critic, parameterized by the trainable parameters \(\psi^{\mathsf c}\). Both critics are trained to approximate their respective \(\mathrm{TD}(\lambda)\) targets using Dreamer-style two-hot categorical representations in the symlog domain. Accordingly, the reward-critic loss is expressed as follows:
\vspace{-5pt}
\begin{equation}
    \mathcal{L}(\psi^\mathsf r)=\mathbb{E}_t\left[-\operatorname{twohot}\!\left(\operatorname{symlog}\!\left(V_{\lambda}^{\mathsf r}[t]\right)\right)^{T}\log {p}_{\psi^\mathsf r}[t]\right],
\end{equation}
while the cost-critic loss is expressed as follows:
\vspace{-5pt}
\begin{equation}
    \mathcal{L}(\psi^\mathsf c)=\mathbb{E}_t\left[-\operatorname{twohot}\!\left(\operatorname{symlog}\!\left(V_{\lambda}^{\mathsf c}[t]\right)\right)^{T}\log {p}_{\psi^\mathsf c}[t]\right],
\end{equation}
where $\operatorname{symlog}(x)=\operatorname{sign}(x)\log\!\left(|x|+1\right)$ \cite{safedreamerV3}.

\subsubsection{Actor Learning}
The actor is optimized to maximize time-averaged EE while reducing minimum-rate constraint violations by leveraging imagined trajectories generated by the world model and future return estimates provided by the reward and cost critics. Based on these estimates, the reward and cost advantages are computed to evaluate the effects of the selected actions on EE and constraint satisfaction, respectively. The reward and cost advantages are respectively defined as $A^ \mathsf r[t]=\frac{V_{\lambda}^{\mathsf r}[t]-v_{\psi^\mathsf r}[t]}{S^\mathsf r}$, $A^\mathsf c[t]=\frac{V_{\lambda}^{\mathsf c}[t]-v_{\psi^\mathsf c}[t]}{S^\mathsf c}$, where \(S^{\mathsf r}\) and \(S^{\mathsf c}\) denote the reward and cost return-normalization scales, respectively. To balance EE maximization and minimum-rate constraint satisfaction, the combined advantage is formulated as $A[t]=\frac{A^\mathsf r[t]-\lambda^\mathsf c A^\mathsf c[t]}{1+\lambda^\mathsf c}$ where \(\lambda^{\mathsf c}\) denotes the effective Lagrange multiplier, which controls the relative importance of safety during actor optimization. The effective Lagrange multiplier is updated as \(\lambda^{\mathsf c}=\max\{0,\lambda^{\mathsf L}+\mu^\mathsf c(J^{\mathsf c}-b)\}\), where $\mu^ \mathsf c$ denotes the penalty coefficient and \(\lambda^{\mathsf L}\leftarrow[\lambda^{\mathsf L}+\mu^\mathsf L(J^\mathsf c-b)]_{0}^{\lambda^{\mathsf{max}}}\) is the Lagrange multiplier iteratively updated based on the observed safety cost, with \(\mu^\mathsf L\) representing the dual step size, and \(\lambda^{\mathsf{max}}\) specifying the maximum allowable multiplier. Furthermore, \([x]_0^{\lambda^{\mathsf{max}}}=\min\{\lambda^{\mathsf{max}},\max\{0,x\}\}\) denotes the clipping operation. Accordingly, the actor loss is expressed as \cite{safedreamerV3}:
\vspace{-5pt}
\begin{equation}
\begin{aligned}
\mathcal{L}(\boldsymbol{\theta})
=-\mathbb{E}_{t}\Big[
\Big(
&\mathrm{sg}\big(A[t]\big)\log \pi_{\boldsymbol{\theta}}
\big(\mathbf{a}[t]\mid\mathbf{h}[t],\mathbf{z}[t]\big)\\
&+\xi\mathcal{H}\big[
\pi_{\boldsymbol{\theta}}(\cdot\mid\mathbf{h}[t],\mathbf{z}[t])
\big]
\Big)
\Big],
\end{aligned}
\label{actor_loss}
\end{equation}
where \(\pi_{\boldsymbol{\theta}}\) denotes the stochastic actor policy parameterized by \(\boldsymbol{\theta}\), \(\mathcal H[\cdot]\) represents the policy entropy, and \(\xi\) is the entropy regularization coefficient. This formulation encourages the actor to maximize long-term EE while penalizing minimum-rate constraint violations.
\begin{figure*}[t!]
\centering
    \begin{subfigure}{0.27\textwidth}
        \includegraphics[width=\linewidth]{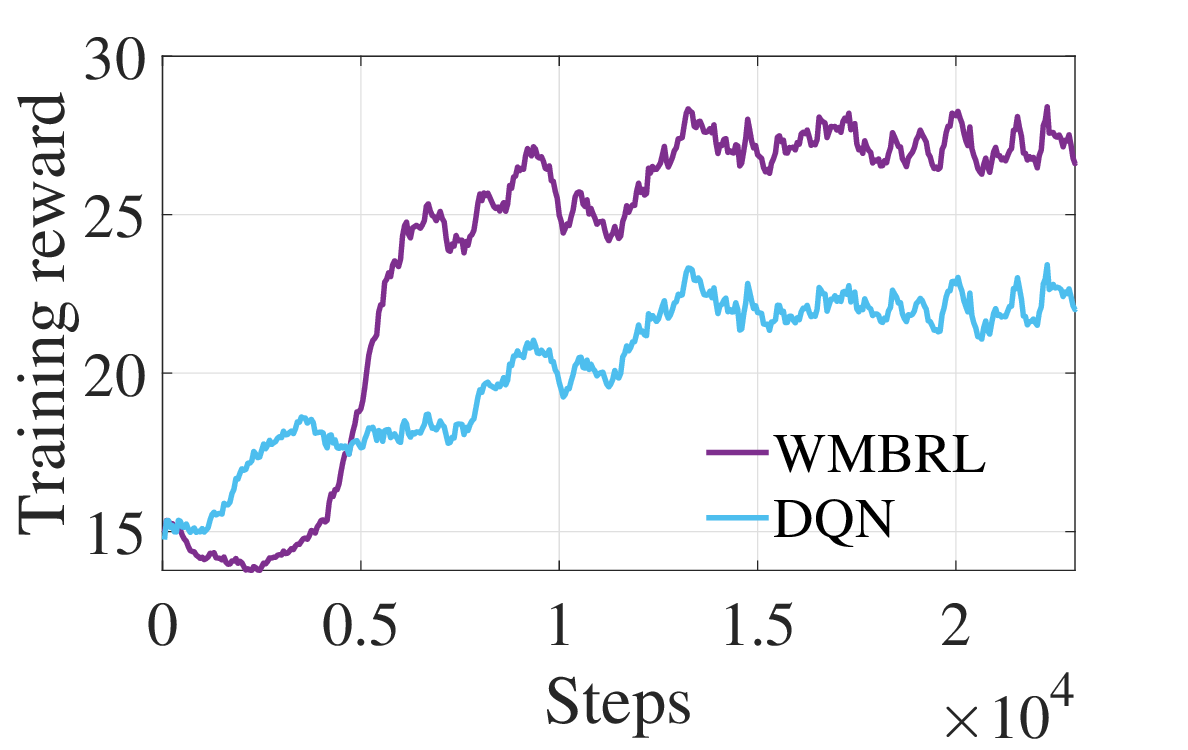}
        \caption{}\label{2a}
    \end{subfigure}
    \begin{subfigure}{0.27\textwidth}
       \includegraphics[width=\linewidth]{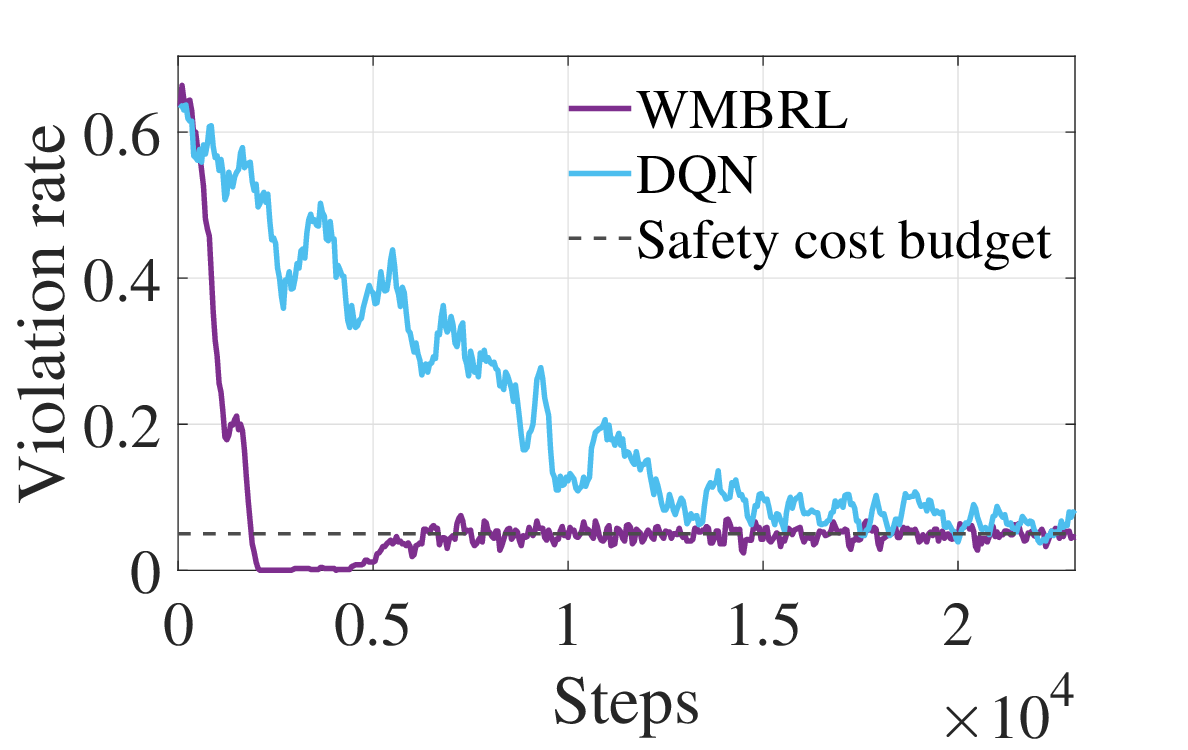}
        \caption{}\label{2b}
    \end{subfigure}
    \begin{subfigure}{0.29\textwidth}
        \includegraphics[width=\linewidth]{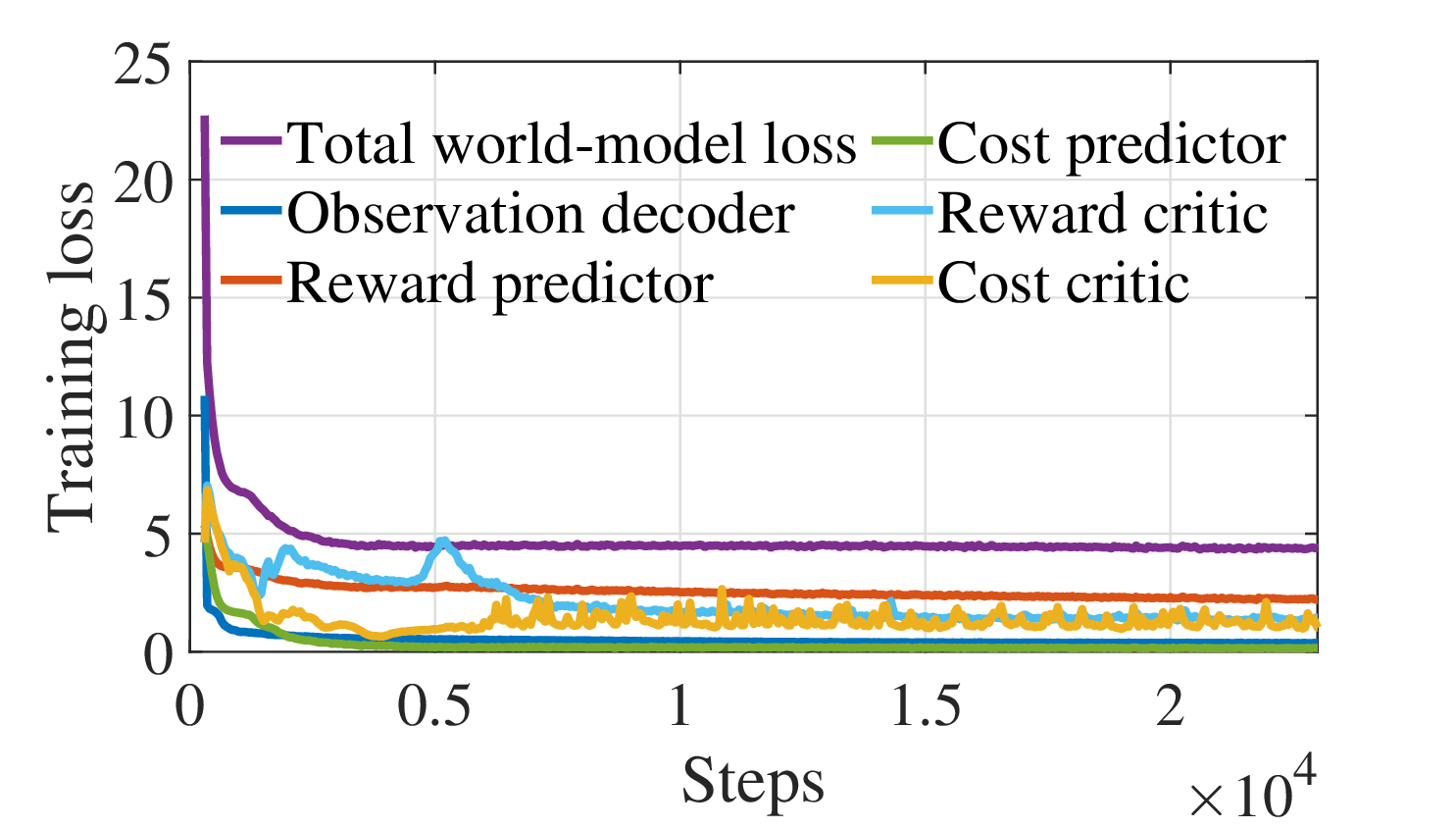}
        \caption{}\label{2c}
    \end{subfigure}
    \begin{subfigure}{0.27\textwidth}
       \includegraphics[width=\linewidth]{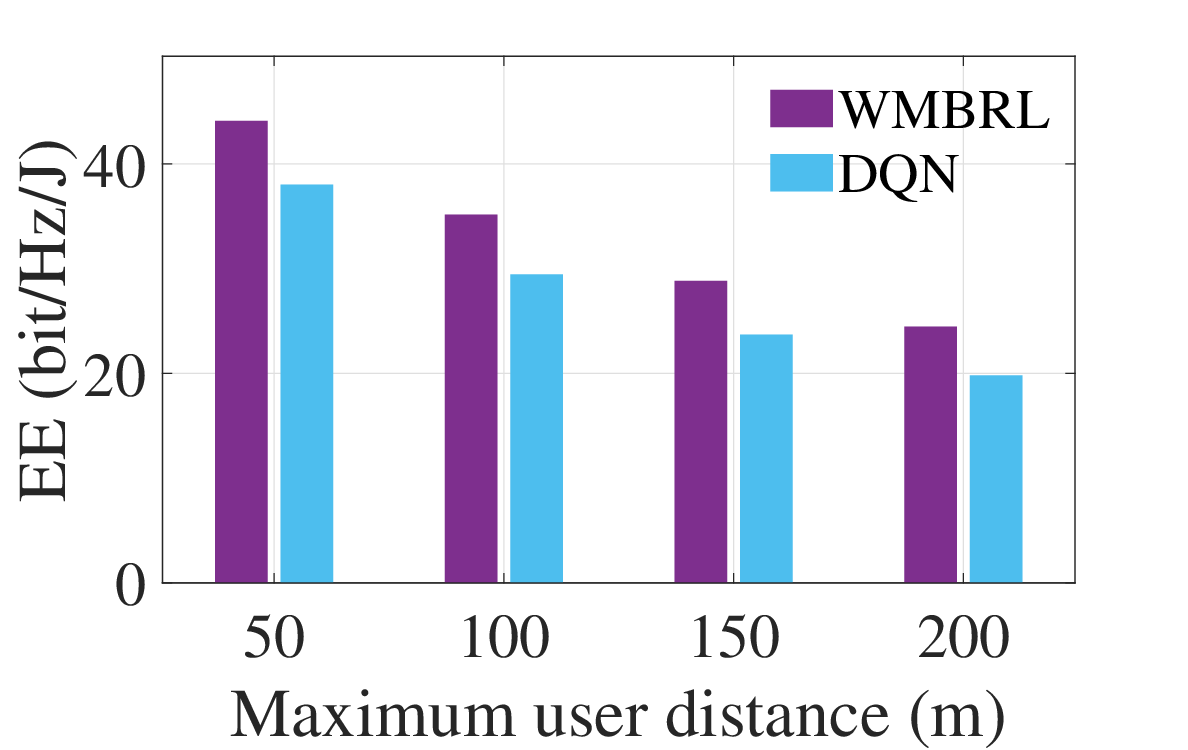}
        \caption{}\label{2e}
    \end{subfigure}
    \begin{subfigure}{0.27\textwidth}
        \includegraphics[width=\linewidth]{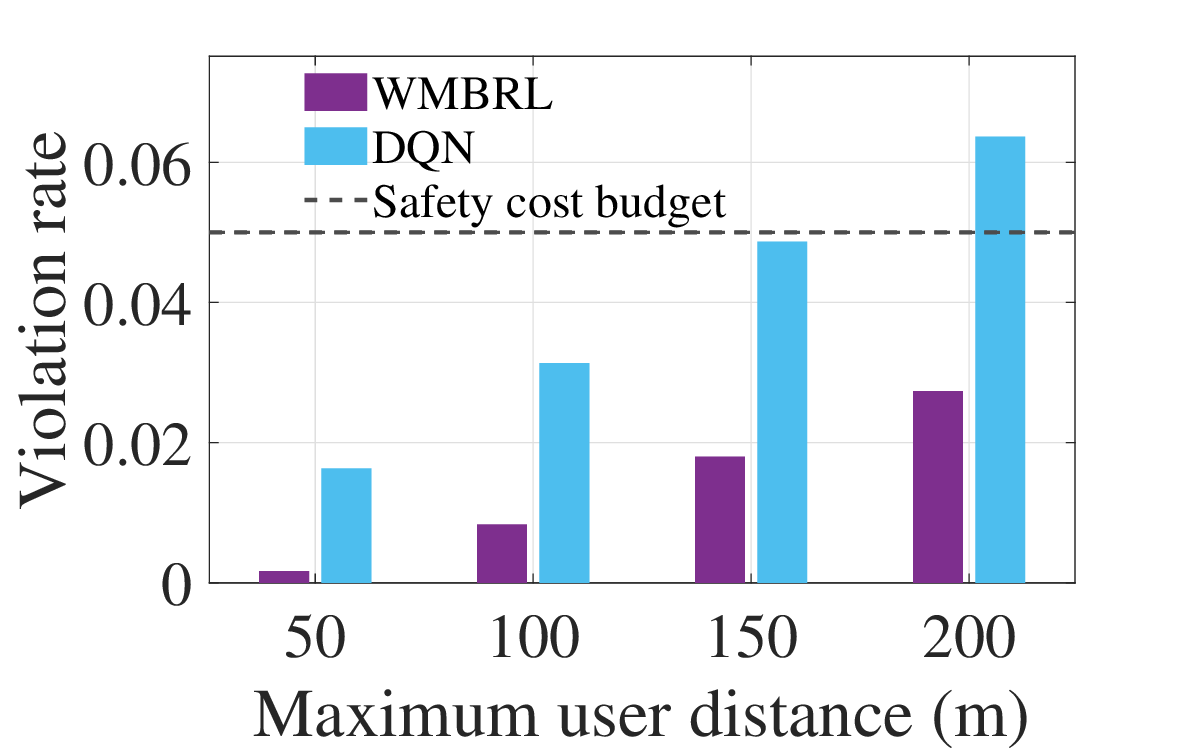}
        \caption{}\label{2f}
    \end{subfigure}
    \vspace{-6pt}
    \caption{Performance evaluation: (a) training reward versus steps, (b) violation rate versus steps, (c) world-model and critic losses versus steps, (d) time-averaged EE versus maximum user distance, and (e) violation rate versus maximum user distance.}
    \vspace{-10pt}
    \label{elem_var}
\end{figure*}

\begin{table}[t]
\centering
\caption{Simulation Parameters}
\label{tab:simulation_parameters}
\renewcommand{\arraystretch}{1.08}
\setlength{\tabcolsep}{5pt}
\begin{tabular}{lc}
\hline
\textbf{Parameter} & \textbf{Value} \\
\hline
\multicolumn{2}{c}{\textit{System Parameters}} \\
\hline
Number of DMA microstrips & $N^{\mathsf d}=4$ \\
Elements per microstrip & $N^{\mathsf e}=6$ \\
Number of users & $K=2$ \\
Carrier frequency & $f^\mathsf c=28$ GHz \\
Time slots per episode & $T=50$ \\
Maximum power budget & $P^{\mathsf{max}}=30$ dBm \\
Circuit power consumption & $P^{\mathsf c}=23$ dBm \\
Minimum radiated power & $P^\mathsf m = 0$ dBm \\
Noise power & $\sigma_k^2=-87$ dBm \\
Minimum rate requirement &
$\Gamma^{\mathsf{min}}=3$ bit/s/Hz \\
Channel correlation & $\rho=0.9$ \\
Waveguide attenuation coefficient & $\alpha=0.6 \,\,\text{m}^{-1}$ \\
Propagation wavenumber & $\beta=827.67 \,\,\text{m}^{-1}$ \\
Minimum user distance & $\psi^{\mathsf{min}}=20$ m \\
\hline
\multicolumn{2}{c}{\textit{WMBRL Parameters}} \\
\hline
RSSM deterministic-state dimension & $256$ \\
RSSM hidden dimension & $256$ \\
Stochastic latent state & $16\times16$ \\
Imagination horizon & $\widetilde{H}=15$ \\
Discount factor & $\gamma=0.97$ \\
TD($\lambda$) parameter & $\lambda=0.95$ \\
Batch size & $16$ \\
World-model learning rate & $10^{-4}$ \\
Actor learning rate & $8\times10^{-5}$ \\
Critic learning rate & $8\times10^{-5}$ \\
Two-hot bins & $255$ \\
Entropy coefficient & $\xi=3\times10^{-4}$ \\
Maximum Lagrange multiplier
    & $\lambda^{\mathsf{max}}=20$ \\
Safety-cost budget & $b=0.05$ \\
\hline
\end{tabular}
\end{table}

\section{Simulation Results}\label{secIV}
In this section, we evaluate the performance of the proposed WMBRL framework through numerical simulations. The system and learning parameters used in the simulations are summarized in Table \ref{tab:simulation_parameters}. The maximum user distance is set to $\psi^{\max}=150$~m in Figs. \ref{2a}--\ref{2c},
while it is varied from 50 to 200~m in Figs. \ref{2e} and \ref{2f}. To demonstrate the effectiveness of the proposed framework, we compare its performance with the DQN baseline that uses Lagrangian-based constraint handling.

Fig. \ref{2a} illustrates the training reward, defined as the objective function of problem \(\mathcal{P}1\) in \eqref{eq:objective_DMA}, of the proposed WMBRL framework and the DQN baseline versus training steps. After convergence, the proposed framework achieves higher training rewards than DQN, demonstrating its effectiveness in improving long-term performance. Fig. \ref{2b} presents the violation rate versus training steps. The proposed framework reduces the violation rate below the prescribed safety-cost budget within approximately \(2\times10^3\) steps and maintains it near the budget thereafter, whereas DQN requires approximately \(1.4\times10^4\) steps to reach this level and exhibits greater fluctuations. These results demonstrate that the proposed framework achieves faster constraint satisfaction and more stable safety performance than DQN.

Fig. \ref{2c} illustrates the convergence behavior of the world-model and critic losses. The total world-model loss decreases rapidly and stabilizes around 4.3, while the observation, reward, and cost prediction losses converge to relatively stable values. Similarly, both critic losses decrease substantially during the initial training stages and subsequently exhibit moderate fluctuations, indicating stable learning behavior.

Fig. \ref{2e} illustrates the impact of the maximum user distance on time-averaged EE. As the maximum distance increases from 50 to 200 m, the EE of both schemes decreases due to increased path loss and reduced channel gains. Nevertheless, the proposed WMBRL framework consistently outperforms DQN across all considered distances, achieving an average improvement of approximately 20\% in time-averaged EE. These results demonstrate the effectiveness of the proposed framework in maintaining superior EE performance under varying user distances.

Fig. \ref{2f} illustrates the impact of the maximum user distance on violation rate. As the maximum user distance increases from 50 to 200 m, both schemes exhibit a higher violation rate due to increased path loss and reduced achievable rates. Nevertheless, WMBRL consistently achieves a lower violation rate across all considered distances, remaining below the prescribed safety budget, whereas DQN exceeds this threshold at 200 m. These results demonstrate that the proposed framework keeps the violation rate below the prescribed safety-cost budget across all considered user distances.

\section{Conclusion}\label{secV}
In this paper, we proposed a WMBRL framework for jointly optimizing digital beamforming and DMA element weights to maximize time-averaged EE under time-varying channel conditions. The framework leveraged learned latent dynamics and imagined trajectories for sample-efficient long-term policy optimization, while employing feasibility-preserving action mappings and a safety-cost mechanism to handle operational constraints. Simulation results demonstrated that the WMBRL achieved an average improvement of approximately 20\% in time-averaged EE over DQN, while exhibiting faster convergence and more reliable constraint satisfaction.

\bibliographystyle{IEEEtran}
\bibliography{bib.bib}

\end{document}